\documentclass[aps,reprint,showpacs,twocolumn]{revtex4-1}
\usepackage{graphicx,color,dcolumn,bm,float,longtable,mathptmx}
\usepackage{amsfonts,amsmath,amssymb}
\usepackage{longtable}
\usepackage{float}

\newcommand{\kk}{\bf k}

\begin{document}

\preprint{ver. 0.0 / \today}
\title{Strain-induced Giant Second-harmonic Generation
  in Monolayered $2H$-MoX$_2$ (X=S,Se,Te)}
\author{S. H. Rhim$^{1,2}$}
\email{sonny@ulsan.ac.kr}
\author{Yong Soo Kim$^{2}$}
\author{A. J. Freeman$^1$}
\affiliation{
  $^1$ Department of Physics and Astronomy, Northwestern University, Evanston,IL, 60208, USA\\
  $^2$ Department of Physics and Energy Harvest Storage Research Center,
  University of Ulsan, Ulsan, Republic of Korea }
\date{\today}

\begin{abstract}
  Dynamic second-order nonlinear susceptibilities,
  $\chi^{(2)}(2\omega,\omega,\omega)\equiv \chi^{(2)}(\omega)$,
  are calculated here within a fully first-principles scheme
  for monolayered molybdenum dichalcogenides,
  $2H$-MoX$_2$ (X=S,Se,Te).
  The absolute values of $\chi^{(2)}(\omega)$ across
  the three chalcogens critically depend on the band gap energies upon uniform strain,
  yielding the highest $\chi^{(2)}(0)\sim$ 140 pm/V
  for MoTe$_2$ in the static limit.
  Under this uniform in-plane stress, $2H$-MoX$_2$ can undergo
  direct-to-indirect transition of band gaps,
  which in turn substantially affects $\chi^{(2)}(\omega)$. 
  The tunability of $\chi^{(2)}(\omega)$ by either compressive or tensile strain
  is demonstrated especially for two important experimental wavelengths,
  1064 nm and 800 nm, where resonantly enhanced non-linear effects
  can be exploited: $\chi^{(2)}$ of MoSe$_2$ and MoTe$_2$
  approach $\sim$800 pm/V with -2\% strain at 1064 nm.
  \end{abstract}
\maketitle

Graphene, two-dimensional material, has attracted great attention
for their intriguing physics such as 
exteremely high mobility\cite{novoselov05_Nature,columbia_2005,geim-novo-nat_mat_2007,neto09:RMP}.
However, their semi-metallic character due to the absence of a band gap,
hinders graphene to replace for silicon based technology.
There have been a lot of research efforts to artificially create
band gap in graphene by introducing defects, ripples, and so forth.

MoS$_2$, another two-dimenionsional material, has also revealed
fascinating proprieties found in graphene:
high mobility and mechanical strength\cite{mak10:prl},
and even superconductivity\cite{ye12:sci}.
Despite many similarities to graphene,
the hetero-atomic constitution of MoS$_2$ 
naturally breaks valley {\em or} sublattice degeneracy.
As a result, the MoS$_2$ band gap is finite,
as large as 1.3 eV\cite{splendiani10:nl,mak10:prl}.
Other materials in the same family,
the so-called transition-metal dichalcogenides (TMDs), {\em or} $2H$-MoX$_2$ (X=S,Se,Te),
possess similar properties due to their equivalent structures.

As graphene has been easily isolated from graphite,
few-layer MoX$_2$ has also been successfully exfoliated from bulk MoX$_2$.
Interestingly, while bulk MoS$_2$ has an indirect band gap,
the isolated monolayer-MoS$_2$ reveals a direct band gap\cite{mak10:prl,splendiani10:nl,qiu13:prl}.
This indirect-to-direct transition of the band gap is not restricted to MoS$_2$:
other TMDs also exhibit the indirect-to-direct transition of the band gap
in their monolayered structure\cite{Yun-12:prb}.

\begin{figure}[b]
 \centering
  \includegraphics[width=0.75\columnwidth]{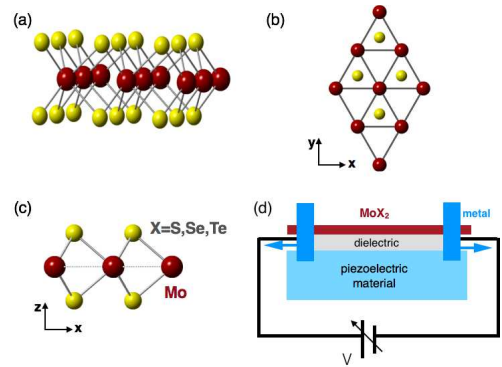}
  \caption{(color online) (a) Structure of the monolayer $2H$-MoX$_2$ (X=S,Se,Te).
    (b) Top- and (c) side-view of the structure of the monolayer of $2H$-MoX$_2$. 
    Red and yellow spheres denote Mo and chalcogen (S,Se,and Te) atoms, respectively.
    (d) A schematics to induce desired strain experimentally. 
  }
  \label{fig:1}
\end{figure}

Bulk MoS$_2$ crystallizes in the hexagonal structure
with centrosymmetric space group $P6_3/mmc$ (No. 194). 
This centrosymmetricity is broken in the monolayer,
which is illustrated in Fig.~\ref{fig:1}(a-c):
One unit layer has one Mo, and two S atoms above and below Mo,
each taking $A$ and $B$ sites, respectively.
The Mo-X bonds are strongly covalent,
and the sandwich layers are weakly coupled by  van der Waals interactions.

Among many promising properties, 
the feasibility of achieving large second-harmonic-generation (SHG)
has recalled intensive attention for an immediate practical application.
In the exfoliated MoS$_2$ film,
a strong second-order non-linear optical properties is observed
in the odd-layer MoS$_2$ as a consequence of broken centro-symmetricity,
which vanishes for the even-layer ones due to the centro-symmetricity\cite{YLi:NL2013}.
To date, several research groups have conducted
measurements on the second-order non-linear optical properties on MoS$_2$ monolayer
\cite{YLi:NL2013,clark14:prb,Malard13:prb,Kumar-2013:prb}. 
However, their values differ by three orders of magnitude from one measurement to another.
The origin of this discrepancy is still under debate\cite{clark14:prb}.
Nevertheless, unarguably large SHG coefficient, reaching close to 100 pm/V,
makes MoS$_2$ very promising for non-linear
optics applications\cite{YLi:NL2013,clark14:prb,Malard13:prb,Kumar-2013:prb,trolle13:arXiv}. 
\begin{figure}[t]
  \centering
  \includegraphics[width=0.8\columnwidth]{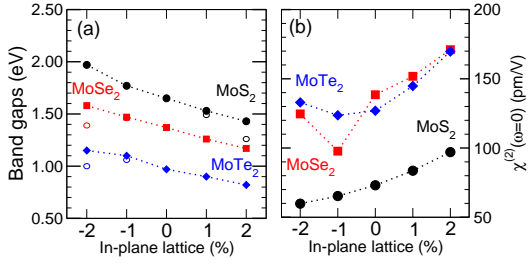}
  \caption{
    (a) Band gaps (in eV) and (b) Static value of second-harmonic generation coefficients,
    $\chi^{(2)}(\omega=0)$ (in pm/V) of 2$H$-MoX$_2$
    (X=S,Se,Te) for different lattice constants.
    Black, red, and blue lines denote those of MoS$_2$, MoSe$_2$, and MoTe$_2$, respectively.
    Open symbols in (a) denote magnitudes of indirect band gaps.}
 \label{fig:2}
\end{figure}

In this letter, using fully first-principles calculations of $2H$-MoX$_2$ (X=S,Se,Te) monolayer,
the electronic structure and the SHG coefficients are investigated.
In-plane strain in a uniform manner alters band gaps,
which in turn changes the SHG coefficients.
An experimental way to realize such strain is proposed in Fig.~\ref{fig:1}(d).
The TMD is placed above piezoelectric/dielectric materials such as PZT,
whose both ends are pinned by metals.
Applied electric field will physically change the shape of the substrate,
which in turn changes lattice constant of TMD. 
Despite qualitative similarities of the electronic structure
among the $2H$-MoX$_2$,
the resultant frequency-dependent SHG coefficients, $\chi^{(2)}(\omega)$,
show some differences in their spectra.
The tunability of $\chi^{(2)}$ by chalcogen atoms, as well as the in-plane strain, will be addressed.

First-principles calculations are carried out using the highly precise
full-potential linearized augmented plane wave (FLAPW) method\cite{wimmer:81,flair}.
The local density approximation (LDA)\cite{hl} is employed
for the exchange-correlation potential.
Spin-orbit coupling is explicitly included in a semi-relativistic way.
Muffin-tin radii of 2.25 (Mo) and 2.00 a.u. (chalcogens) are chosen.
Cutoffs for wave function and potential representations are 14.06 and 144 Ry., respectively.
The angular momentum expansion of charge density and potential 
inside the MT sphere are done for $\ell \le 8$.
Summation in the Brillouin zone is performed
using Monkhorst-Pack scheme with a 45$\times$45$\times$15 {\em k} point mesh,
which includes high-symmetric $K$ and $M$ points.
To reduce artificial layer-layer interactions,
adjacent layers are separated by 20~\AA. 
We employed the experimental lattice constants for the present work,
$a$=3.160~\AA~(MoS$_2$), 3.299~\AA~(MoSe$_2$), and 3.522~\AA~(MoTe$_2$), respectively.
Internal coordinates has been fully relaxed with force criteria 1$\times$10$^{-3}$ eV/\AA.
For simplicity, band gap correction beyond LDA is not employed
and the exciton effects are ignored. 

For calculations of the second-harmonic generation (SHG) coefficients,
we employ the formalism by Duan {\em et al.}\cite{duan99:prb,duan99-2:prb},
which is an extension of method by Sipe and Ghahramani\cite{sipe93:prb} and Aversa and Sipe\cite{Sipe95:prb}.
This formalism has been very successful in other works
\cite{duan99:prb,duan99-2:prb,lambrecht00:pss,cheiw12:prb,jhsong09:prb,stampos15:jacs},
where the imaginary part $\chi^{(2)}_{abc}$ is calculated as
%
\begin{widetext}
\begin{eqnarray}
  \label{eq:1}
 \chi^{(2)}_{abc}(\omega) &=&
-\frac{1}{2}\int_{{\rm BZ}}\frac{d\kk}{4\pi^3}\sum_{ijk}{\rm Im}\left[ p^{a}_{ij}\{ p^{b}_{jl} p^{c}_{li}\}\right]
\Big(
\frac{16}{\omega^3_{jl}(2\omega - \omega_{ji})}
[
\frac{f_{il}}{\omega_{jl} - 2\omega_{li}} + \frac{f_{jl}}{ \omega_{ji} - 2\omega_{jl}}
] \\ \nonumber
&+&
\frac{f_{il}}{ \omega^{3}_{li} \left( 2\omega_{li}-\omega_{ji}\right) \left(\omega-\omega_{li}\right) }
+
\frac{f_{jl}}{\omega^{3}_{jl} \left(2\omega_{jl}-\omega_{ji}\right)\left(\omega-\omega_{jl}\right)}
\Big)
\end{eqnarray}
\end{widetext}
where $(a,b,c)$ represent the cartesian $(x,y,z)$, 
and $i,j$ stands for the valence and the conduction band, respectively,
and $l\ne i,j$ denotes the intermediate virtual state-
either the virtual electron or hole state.
$\{p^b_{jl}p^c_{li}\}=\frac{1}{2}\left(p^b_{jl}p^c_{li}+p^b_{li}p^c_{jl}\right)$,
where $p_{ij}^{a} = \langle i | {\bf p}_{a}| j \rangle$ is momentum matrix element
with suppressed wave vectors for the Bloch state indices.
A $45\times45\times15$ {\em k} mesh is also used for summation
in Eq.~(\ref{eq:1}) using the special {\em k} point scheme. 
The corresponding real part of $\chi^{(2)}_{abc}(\omega)$ is obtained
by the Kramers-Kronig transformation of the calculated imaginary part.
Due to the hexagonal symmetry in monolayered $2H$-MoX$_2$,
there is only one non-vanishing component in the $\chi^{(2)}$ tensor in $D_{3h}$ symmetry:
$\chi^{(2)}\equiv \chi^{(2)}_{xxy}=\chi^{(2)}_{yxx}=-\chi^{(2)}_{yyy}$\cite{SYMM_commemnt}.

The calculated band gaps of $2H$-MoX$_2$ are shown in Fig.~\ref{fig:2}(a),
where the direct (indirect) band gaps are shown in filled (empty) symbols.
Upon pressure {\em or} lattice strain,
band gaps decrease (increase) as the lattice expands (contracts).
Noteworthy, there are indirect band gaps 
for expanded (MoS$_2$) and contracted lattices (MoSe$_2$ and MoTe$_2$).
This feature in connection to the band structure will be discussed later.
Before we proceed, we emphasize here that $\pm$2\% strain is feasible in terms of energetics point of view.
[See Supplemental material at [http://..] for discussion in Sec.I]\cite{SM}


\begin{figure}[ht]
  \centering
  \includegraphics[width=1.\columnwidth]{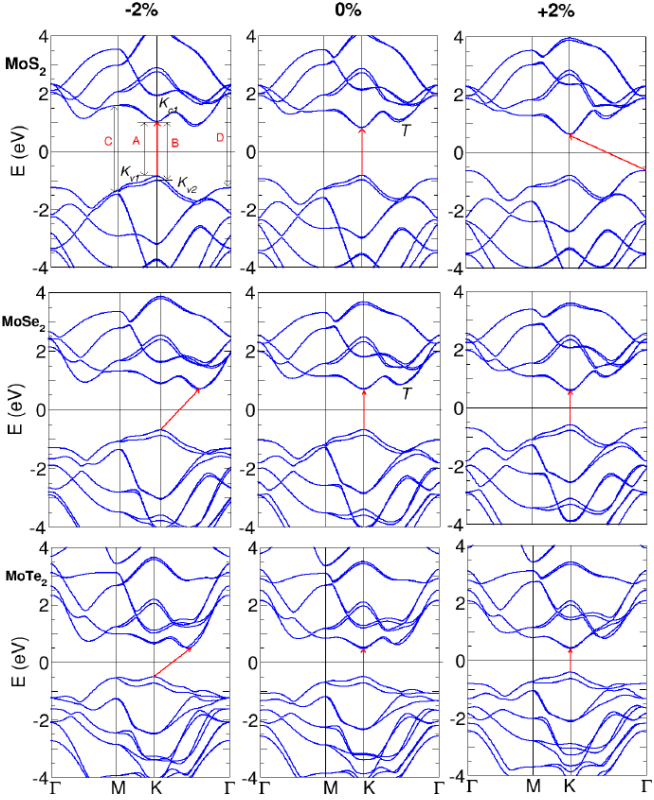}
  \caption{Band structure of monolayer 2H-MoX$_2$ (X=S,Se,and Te) 
    for different lattice strains:
     -2, 0, and +2\%     with respect to the experimental lattice constant.
     Red arrow in each panel indicates whether band gaps either direct or indirect.}
  \label{fig:3}
\end{figure}
Band gaps agree well with experiment.
This is accidental acknowledging LDA band gap underestimation
implying exciton binding energy is large.
Other first-principles calculations utilizing many-body effect have revealed
that exciton binding energies are 1.00 eV and 0.62 eV
for MoS$_2$\cite{qiu13:prl} and MoSe$_2$\cite{ugeda14:nmat}, respectively.
That for MoTe$_2$ is not available at this point.
In this work, for simplicity excitons are not fully taken into account.
Neglect of excitons in our work will not alter physics qualitatively.

The SHG coefficients for the static limit, $\chi^{(2)}(\omega=0)$ [$\equiv \chi^{(2)}(0)$],
as a function of different lattice constants are plotted in Fig.~\ref{fig:2}(b).
$\chi^{(2)}(0)$ of MoS$_2$ increases monotonically as the lattice expands
due to the smaller band gaps, reaching 97.04 pm/V for a +2\% change.
For MoSe$_2$ and MoTe$_2$, $\chi^{(2)}(0)$ also increases as the lattice expands.
However, both cases have dips at -1\% lattice constants.
This is due to smaller band gap of -1\% lattice with respect to -2\%.
Moreover, indirect band gaps of -2\% lattice are much smaller
than the direct ones of -1\% lattice by 0.2 eV.
Above all, the enhancement of $\chi^{(2)}(0)$ in most cases is well understood:
lattice expansion decreases the band gap, which in turn increases $\chi^{(2)}(0)$.
Interestingly, $\chi^{(2)}(0)$ of MoS$_2$ reaches almost 100 pm/V,
whereas those of MoSe$_2$ and MoTe$_2$ exceed 100 pm/V with the maximum value as high as 170 pm/V.

For a more quantitative analysis, the band structures of $2H$-MoX$_2$ 
are plotted in Fig.~\ref{fig:3}.
[See Supplemental material at [http://..] for band plots of other strains in Fig.S1]\cite{SM}
Important transitions from the valence band maximum (VBM) to the conduction band minimum (CBM) are labeled.
$A$ denotes the transition from VBM ($K_{{\rm v}1}$) to CBM ($K_{{\rm c}1}$) at the $K$ point,
$B$ denotes from the spin-orbit split states ($K_{{\rm v}2}$) to $K_{{\rm c}1}$,
$C$ and $D$ are transitions at $M$ and $\Gamma$, respectively.
Near the midpoint between $K$-$\Gamma$ in the valence band is labelled as $T_{{\rm v}}$.
Red arrows connect VBM to CBM:

As mentioned previously,
indirect band gaps are shown for expanded (contracted) lattices of MoS$_2$ (MoSe$_2$ and MoTe$_2$).
In all cases, direct band gaps occur at $K$.
On the other hand, indirect gaps of MoS$_2$ occur from the $\Gamma_{{\rm v}1}$ to $K_{{\rm c}1}$;
those in MoSe$_2$ and MoTe$_2$ occur 
from $T_{{\rm v}}$ to $K_{{\rm c}1}$.
SOC splittings of MoS$_2$, MoSe$_2$, and MoTe$_2$ are of 151, 190, and 229 meV, respectively,
which are energy differences of $K_{{\rm v}1}$ and $K_{{\rm v}2}$.
As shown in previous studies\cite{Yun-12:prb,Han-11:prb},
VBM and CBM consist mainly of Mo $d$ states with a weak contribution of chalcogen $p$ states.
In $D_{3h}$ symmetry, $d$ states have three irreducible representations:
$xz/yz$, $xy/x^2$-$y^2$ and $z^2$.
More specifically, at the $K$ point, the VBM and CBM are predominantly derived
from $xy/x^2$-$y^2$ and $z^2$ states, respectively.
On the other hand, $z^2$ ($xz/yz$) dominates the VBM (CBM) near $\Gamma$ point.
At $T$, midpoint of $K$-$\Gamma$,
 while the CBM is mainly of $x^2$-$y^2/xy$ state,
the VBM is of a mixture of $x^2$-$y^2/xy$ and $xz/yz$.
In MoSe$_2$ and MoTe$_2$, on the other hand,
while dominant states in $K_{{\rm v}1,2}$, $K_{{\rm c}}$, and $\Gamma_{{\rm c,v}}$ are similar to MoS$_2$,
the valence band of the midpoint, $T_{{\rm v}}$, consists of a mixture of $xz/yz$ and $z^2$.
We note that indirect gaps in MoS$_2$ occur such that
$\Gamma_{{\rm v}}$ shifts upward with lattice contraction,
while in MoSe$_2$ and MoTe$_2$ $T_{{\rm v}}$ moves upward with lattice expansion.
Since the $d$ contribution of $\Gamma_{{\rm c}}$ in MoS$_2$
and of $T_{{\rm v}}$ of MoSe$_2$ and MoTe$_2$ are both $xz/yz$,
contraction of lattice in MoS$_2$ acts similarly as expansion in MoSe$_2$ and MoTe$_2$.
\begin{figure}[ht]
  \centering
  \includegraphics[width=1.\columnwidth]{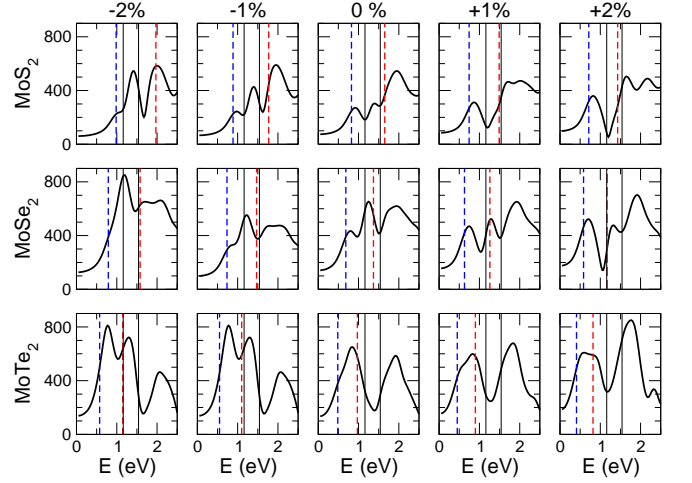}
  \caption{Second harmonic coefficient (in pm/V) of $2H$-MoX$_2$ (X=S,Se, and Te)
    for different lattice strains:
     -2, -1, 0, +1, and +2\%
    with respect to the experimental lattice constant.
    Blue and red dotted vertical lines denote half of the band gaps ($H$) and band gaps ($E$).
    Two vertical lines indicate infrared frequency range,
    1064 nm (1.16 eV) and 800 nm (1.54 eV). }
 \label{fig:4}
\end{figure}

The frequency-dependent dispersion of the SHG coefficients, $\chi^{(2)}(\omega)$,
are plotted in Fig.~\ref{fig:3} for different strains.
The frequency window is chosen from 0 to 2.5 eV,
which includes the important frequency range (IR) of 1064 nm (1.16 eV) and 800 nm (1.54 eV).
[See Supplemental material at[http://..]\cite{SM}
to refer frequency-dependent $\chi^{(2)}(\omega)$ for a wider range of frequencies in Fig.S2.]

For the MoS$_2$, in the absence of strain,
the large peak around 2.0 eV reaches 600 pm/V
just above the band edge around 1.0 eV.
Below the band edge ($E$),
two smaller peaks appear at half of the band gap ($H$)
and 0.2 eV below the band gap.
From the band analysis,
also in qualitative agreement
with 
other work
on the SHG coefficients of MoS$_2$\cite{clark14:prb,trolle13:arXiv},
these two peaks are from $A/2$, $B/2$ and $C/2$, $D/2$, respectively,
where the division by two indicates
half of the energy values of the corresponding transitions.
The enhancement of $\chi^{(2)}(\omega)$ is apparent for +1 and +2\% lattice change
at a frequency much larger than 2.0 eV.
On the other hand, large peaks also occur at lower frequencies for contracted lattices.
In particular, the peak around the IR frequency range is prominent: $\chi^{(2)}\sim$ 600 pm/V.

Four distinct peaks are apparent for $\omega < 4$ eV
in the case of MoSe$_2$ with zero strain.
Two peaks at $H$ and at 3 eV are small,
and the peak just below $E$ is pronounced,
while that at 2 eV is rather broadened.
For +1\% strain, the first peak at $H$ is sharpened,
while at $E$ it is still sharp  but reduced.
On the other hand, for +2\% strain, 
there is a dip between peaks at $H$ and $E$,
where that at the band edge becomes broadened 
with third peak gets more pronounced.
The change of spectrum between these two stretched strains
originates from the change of the bands:
the top valence bands remains almost unaltered,
while the conduction band changes, though a little bit, between $\Gamma$-$M$ and $K$-$\Gamma$.
For the contracted lattices, the direct band gap becomes indirect,
where the CBM and the $K_{{\rm c}1}$ are very close in energy,
hence the two peaks below $E$ tend to merge.

For the zero strain of MoTe$_2$, two peaks are clearly noticeable,
where the first peak is just below $E$,
while the second peak is well above $E$ near 2 eV.
For positive strains, two peaks slightly shift to lower energies.
More specifically, for +2\% strain, the first peak is a little smoothened between $H$ and $E$,
whereas the second peak just below 2 eV is more enhanced.
On the other hand, for negative strains,
the first peak is split into two - one above $E$,
the other near $H$ with enhancement with respect to the zero strain case.
We note that $\chi^{(2)}(\omega)$ within IR frequency range approaches 800 pm/V
for both MoSe$_2$ and MoTe$_2$ when lattice is contracted by 2\% (-2\% strain) .

In conclusion,
we predict using first-principles calculations 
giant non-linear second-harmonic susceptibilities
of the molybdenum dichalcogenides, $2H$-MoX$_2$ (X=S,Se,Te).
We have found that band gaps can be altered by lattice strains:
Expansion (contraction) of the lattice leads to a decrease (increase) of the band gaps.
As a result, static values of the second-harmonic susceptibilities,
$\chi^{(2)}(0)$, can be altered by lattice strains.
Frequency-dependent SHG coefficients, $\chi^{(2)}(\omega)$,
for different strains are also investigated,
where the peak structure strongly depends on little changes in the electronic structure.
Thus, we have shown that large values of $\chi^{(2)}$ are accessible
by applying lattice strains
for IR frequency range.
In our work, the exciton spectra are ignored.
If the formation of excitons are fully taken into account,
$\chi^{(2)}(\omega)$ should be red-shifted by amount of exciton binding energies.
Moreover, if those excitons are robust in optical spectra such as absorption coefficients and dielectric function,
there will be additional peak at the lower end of the $\chi^{(2)}(\omega)$.

  SHR is grateful to J.I. Jang and Y.D. Jho for fruitful discussions.
  This work supported by the Department of Energy (DE-FG02-88ER45382)
  and Supercomputing time grant by NERSC.
  Work at Ulsan is supported by Priority Research Centers Program (NRF-2009-0093818) 
  and the Basic Science Research Program
  through the National Research Foundation funded
  by the Ministry of Education of Korea, ICT and Future planning (NRF-2015R1A2A2A01003621).


\begin{thebibliography}{29}
\bibitem[{SYM()}]{SYMM_commemnt}
\bibinfo{note}{The exact components may change depending on the choice of
  crystal orientation. We choose the axis as shown in Fig.~\ref{fig:1}.}
\end{thebibliography}
\end{document}